\documentclass[preprint,showpacs,preprintnumbers,amsmath,amssymb]{revtex4} 
\usepackage{graphicx}% Include figure files
\usepackage{dcolumn}% Align table columns on decimal point
\usepackage{bm}% bold math

\begin{document}

\title{\bf Anomalous spin-orbit effects in a strained InGaAs/InP quantum well structure}

\author{S.A. Studenikin\footnote{sergei.studenikin@nrc.ca}, P.T. Coleridge, P. Poole, and A. Sachrajda}
\affiliation{Institute for Microstructural Sciences, National
Research Council of Canada, Ottawa, Ontario, K1A OR6, Canada }
\date{27 January, 2003 }% It is always \today, today,
             %  but any date may be explicitly specified
\begin{abstract}

 There currently is a large effort to explore spin-orbit effects in semiconductor structures with the ultimate goal of manipulating electron spins with gates.  
A search for materials with large spin-orbit coupling is therefore important.
We report results of a study of spin-orbit effects in a strained InGaAs/InP quantum well.   
The spin-orbit relaxation time, determined from the weak antilocalization effect, was found to depend non-monotonically on gate voltage.  The spin orbit scattering rate had a maximum value of 
$5\times 10^{10}s^{-1}$ at an electron density of $n=3\times 10^{15} m^{-2}$.
The scattering rate decreased from this for both increasing and decreasing densities.
The smallest measured value was approximately $10^9 s^{-1}$ at an electron concentration of $n=6\times 10^{15} m^{-2}$. 
This behavior could not be explained  by either the Rashba or the bulk Dresselhaus mechanisms but is attributed to asymmetry or strain effects at dissimilar quantum well interfaces.

\end{abstract}



\maketitle
\newpage

\subsection{Introduction}

In $A_3B_5$ semiconductors containing heavy metal elements, such as indium,
spin-orbit effects are large because of the increased coupling between valence and conduction bands associated with strong relativistic effects.  In bulk materials, without magnetic impurities, the only important mechanism producing spin-orbit coupling is the bulk inversion asymmetry \cite{OptOrientation, Kacharovskii86}.
In contrast, in semiconductor heterostructures, the electron spin-orbit interaction can also be controlled by modification of the subband structure using gate voltages \cite{Miller2003}, strain \cite{Ekenberg2001}, or selective doping \cite{Nitta2002}.

The role of spin-orbit effects in semiconductors is gaining a  significant attention because of recent interest in the emerging fields of spintronics and quantum computation \cite{DiVincenzo1997, Wolf2001}.   Key issues are the injection and detection of spin-polarized electrons, controlling and manipulating single spins, and the design and experimental realization of novel spintronic devices such as spin-transistors, logic elements, and memory. 
One obvious way to inject polarized electrons is to use ferromagnetic contacts \cite{Johnson2002, Hanbicki2002}.  It is also possible to exploit the spin-polarized edge states in lateral quantum dot devices subject to a magnetic field \cite{Andy2001}.  
Another, more challenging approach, is to create a spin-polarized current by employing spin-orbit coupling \cite{Koga2002}.
With the ultimate intention of learning how to manipulate and measure electron spins locally by using gates to modify the spin-orbit interaction, there is a vital interest in searching for semiconductor materials and structures where the electron spin-orbit interaction is large and highly sensitive to gate voltages.

One method which gives information about the spin-orbit coupling is the weak anti-localization (WAL) effect. This is a quantum interference correction to the conductivity which appears as an abnormal positive magnetoresistance in very low magnetic fields, preceding the more usual negative magnetoresistance associated with weak localization \cite{Altshuler85}. In metals WAL was thoroughly studied and understood in the 1980's \cite{Bergman84}. In semiconductors the situation is more complex because new mechanisms involving spin orbit effects come into play, such as bulk nonsymmetry, asymmetry at heterointerfaces and in quantum wells, and two-dimensional quantum confinement. \cite{OptOrientation} 

In this work the WAL effect is used to study spin relaxation due to the spin-orbit interaction in a strained InGaAs/InP quantum well (QW) structure. Compared with an isomorphous (lattice-matched) structure \cite{ourWALpaper}, the strained QW structure showed a larger sensitivity to the gate voltage.  
In addition an anomalous, non-monotonic, dependence of the spin-orbit time constant on electron concentration was observed. This can not be explained by the conventional bulk inversion and/or Rashba mechanisms and suggests that the existing theoretical understanding of spin orbit effects in transport phenomena in semiconductor structures needs to be improved.

\subsection{Experimental}

The QW structure studied here was grown on a (100) InP semi-insulated substrate and consisted of the following layers (measured up from the substrate): 450 nm of undoped InP buffer layer, 10 nm In$_{x}$Ga$_{1-x}$As ({\it x}=0.76) quantum well,  13 nm undoped InP spacer layer, followed by 13 nm InP doped layer ($N_d=4\times10^{23} m^{-3}$) and a 13 nm undoped cap layer.   The indium content in the quantum well was higher than that of InGaAs lattice-matched to InP ({\it x}=0.53) so the QW was compressively strained. 

Standard optical photolithography and wet etching was used to form a 0.2 mm wide Hall bar with 0.4 mm separation between adjacent potential contacts. A 40 nm SiO$_2$ dielectric layer and a gold gate were deposited on top of this. Measurements were performed in a He3 system; experimental details are given in Ref.\onlinecite{ourWALpaper}. 
 
Figure 1 shows the electron concentration determined from Shubnikov-de Haas oscillations and the Hall mobility ($\mu$). These transport properties are very similar to those observed earlier in the isomorphous lattice-matched sample studied previously \cite{ourWALpaper}. In particular, the concentration varies linearly with gate voltage while the mobility has a somewhat slower dependence.

The WAL was used to investigate the spin-orbit scattering action to the conductivity appears as a non-monotonic dependence of the magnetoresistance at very low magnetic fields, $\mu B \ll 1$. An initial positive magnetoresistance is followed by the more usual negative term.    
In the theoretical description of the interference corrections characteristic values of the conductance and magnetic field appear, analogous to the Bohr radius and energy in the theory of excitons. 
It is therefore convenient to plot the conductance (inverse resistance) in units of the quantum conductivity $G_0=(e^2/\pi h)$ and the magnetic field normalized by a characteristic field $B_{tr}$ given by $\hbar/4eD \tau$ where $D=v_F^2\tau /2=\hbar^2\pi n\mu/m^*e$
is the diffusion coefficient of the two dimensional electrons and $\tau$ the transport scattering time.  $B_{tr}$ depends on both electron concentration and mobility and therefore has a stronger gate voltage dependence than the density (see insert to Fig.1), changing by more than an order of magnitude over the range of $V_g$ used in the experiment. 
It should be further noted that because it is desirable to eliminate the irrelevant classical Lorentz term in the magnetoconductivity 
$\sigma_{xx}(B) = \sigma_0/(1+\mu^2 B^2)$ it is also  convenient to plot the inverse magnetoresistance $\Delta (1/\rho_{xx})= 1/\rho_{xx}-1/\rho_0$ rather than $\Delta \sigma=\sigma_{xx} - \sigma_0$, where $\sigma_0$ is the zero field value and $\rho_0 = 1/\sigma_0$. This procedure, which would produce zero in the absence of quantum interference corrections, removes the Lorentz term.

Figure 2 shows experimental traces plotted in this way for different gate voltages. The narrow peak around zero magnetic field is the WAL effect which appears when the spin-orbit scattering rate is comparable to or large than the inverse phase breaking time $1/\tau_{\varphi}$. It is clear, that in this sample, the WAL effect shows a non-monotonic dependence on gate voltage reaching a maximum around $V_g$=-0.3 and decreasing for both large positive and large negative voltages.  Such behavior is unusual and is descussed below.  

To extract the phase-breaking and spin-orbit scattering times we attempted to fit the experiment with the theoretical expression, derived for arbitrary magnetic fields \cite{knap96}:

\begin{equation}
     \Delta\sigma(B) = -(e^2/\pi h) [ F(x,\beta_{s1}) +     
\frac{1}{2} F(x,\beta_{s2}) - \frac{1}{2} F(x,\beta_{\varphi})]
\label{eq1}
\end{equation}

 where
\begin{equation}
     B_{tr} = \frac{\hbar}{4eD\tau} , \; \;
     B_{so} = \frac{\hbar}{4eD\tau_{so}} \;\;\mbox{ and } \;\; 
    B_{\varphi} = \frac{\hbar}{4eD\tau_{\varphi}} 
\label{eq2}
\end{equation}

\begin{displaymath}
x = \frac{B}{B_{tr}} = \frac{4eBD\tau}{\hbar}, \;\;   
\beta_{\varphi} = \frac{\tau}{\tau_{\varphi}}, \;\;
\beta_{so} = \frac{\tau}{\tau_{so}}, \;\;
\beta_{s1} = \beta_{\varphi} + \beta_{so},\;\;
\beta_{s2} = \beta_{\varphi} + 2\beta_{so},
\end{displaymath}

with D the diffusion coefficient and $\tau$, $\tau_{so}$ and $\tau_{\varphi}$  respectively the elastic scattering time, the
spin-orbit relaxation time and the phase-breaking time. The function $F(x,\beta_i)$,  defined in Refs. \onlinecite{knap96,ourWALpaper}, describes the interference contributions from the three triplet and one singlet diffusion channels.
This equation, derived for an arbitrary magnetic field,  reduces to the well known  expression given by Hikami, Larkin and Nagaoka (HLN) \cite{hikami80} in the limit of small magnetic fields $B \ll B_{tr}$. For the fitting procedure $B_{tr}$ is known so there are two adjustable parameters
$\beta_{\varphi}$ and $\beta_{so}$ or equivalently ($\tau_{\varphi}$ and $\tau_{so}$). 
One more remark should be added here: to fully describe the WAL effect requires a spin-dependent vector potential with a three dimensional character \cite{lyandageller98,pikus95, iordanskii94}. 
The two major spin-orbit relaxation mechanisms (Dresselhaus and Rashba) are not additive so in general more complicated expressions, with additional fitting parameters should be used to describe the experiments. If, however, one mechanism dominates a single scalar parameter $\tau_{so}$ suffices which can then be treated on the same footing as $\tau_\varphi$.

The thin solid lines in Fig. 2 are theoretical fits using this Eqn. 1. Details of the fitting procedure are described in Ref. \onlinecite{ourWALpaper}. Although this theory should be valid for
arbitrary  magnetic fields, it was impossible to obtain a good fit over the whole magnetic field range for any of the data. Fitting the central part (small B) resulted in large deviations at higher fields and {\it vice versa}.  This is a common problem in semiconductors encountered by many authors, e.g. 
Ref.\onlinecite{knap96}. A similar large discrepancy between theory and experiment observed in the isomorphous InGaAs QW sample \cite{ourWALpaper} could only be reconciled by introducing an additional,  empirical, scale parameter of order two.  It was argued there that one reason for the discrepancy might be the fact that the spin orbit scattering time was comparable to the transport relaxation time but this is not the case here: for all the curves shown in Fig.2 
$\beta_{so} = \tau /\tau_{so} \leq 0.1$ satisfying the condition $\beta_{so} \ll 1$.    

Given the large discrepancy between experiment and the theoretical description of the WAL effect (which appears to be a general property of high mobility 2DEG semiconductors in high magnetic fields) it is not immediately obvious how to extract values of the spin orbit relaxation time.   Further theoretical effort is needed to fix this problem.   However, the amplitude of the WAL is clearly affected by $\tau_{so}$ and we have chosen to fit it using the low field part of the data where the turnover from a negative to a positive magnetoconductivity is sensitively dependent on $\tau_{so}$. This approach has the advantage that it is also consistent with the procedure commonly used in the literature whereby the low-field peak is fitted to the HLN expression \cite{Nitta2002, hikami80, iordanskii94, dresselhaus92, chen93} with the implicit understanding that deviations at higher fields are to be expected because of inadequacies in the theory.

The fitted theoretical curves shown on Fig. 2 are plotted well beyond the range of the fit to emphasize the unexpected discrepancy between theory and experiment and to stimulate the attention of theorists.

\subsection{Discussion}

  Although the appearance of the WAL effect requires strong spin orbit scattering, the curvature of the WAL peak at B=0 and the characteristic width are in fact not determined by $\tau_{so}$ but rather by the phase breaking time $\tau_{\varphi}$ \cite{ourWALpaper, Ekaterinburg} . It is the amplitude of the WAL, in particular the crossover from WAL to weak localizing behavior, that is determined by the ratio $\tau_{\varphi}/\tau_{so}$.

From Fig. 2 it can seen that the central part of the WAL peak at  B=0  changes little for different $V_g$ and indeed, all the fits gave the same value for parameter 
$\beta_{\varphi}=0.010 \pm 0.001$ (for curve at $V_g=+0.1$, where the WAL had vanished,  $\beta_{\varphi}$ was set to 0.010 and only $\beta_{so}$ fitted).  
The WAL peak is narrow because its width is
determined not by $\beta_{so}$ but rather by $\beta_{\varphi}$ which can be very small in high mobility samples.  Without spin-orbit scattering 
a weak localization peak would appear with the same width but of opposite sign.

The phase breaking time extracted from the fits to the data is plotted in Fig. 3 compared with the predictions of electron-electron scattering calculated from a Fermi liquid model  \cite{Altshuler85,altshuler82}:

\begin{equation}
     \frac{1}{\tau_{\varphi}}=\frac{k_B T}{\hbar}\frac{\pi
G_0}{\sigma_0}ln\left(\frac{\sigma_0}{2\pi G_0}\right)     
\label{eq3}
\end{equation}
with $G_0=e^2/(\pi h)$, and $k_BT\tau/\hbar \ll 1$.  The experimental values of $\tau_{\varphi}$ shown in Fig.3 are all smaller than predicted. In the literature an empirical coefficient of order 2 is often introduced to bring the experimental data into better agreement with Eq. \ref{eq3}  \cite{polyanskaya97, minkov01} but the discrepancy is larger than this.  While this model generally works well in metals, where Fermi-energy is large and the electron gas can be considered as being very uniform \cite{Altshuler85,Bergman84} deviations appear at low temperatures, and  {\it a fortiori} in semiconductors. The phase breaking time is almost universally observed to saturate as the temperature is lowered.  For all the data shown in Fig.3 the temperature was sufficiently low that this saturation had occurred. That is, the absence of any significant gate voltage dependence in $\tau_{\varphi}$ reflects the temperature saturation rather than an intrinsic insensitivity to electron concentration. The saturation implies that there exist additional phase-breaking mechanisms, the analysis of which is not the topic of this paper.

Figure 4 shows the spin-orbit scattering rate determined from the fits to the data in Fig. 2. It is evident that $\tau_{so}$ is a non-monotonic function of the electron concentration. 
In 2DEG systems the two major spin-orbit scattering mechanisms identified in the literature are the Dresselhaus term, associated with the bulk zinc-blend crystal inversion asymmetry, and the Rashba term associated with built-in electric fields \cite{pikus96}. To identify which mechanism dominates here it is helpful to consider the dependence on electron concentration of $B_{so}=\hbar/(4eD\tau_{so})$ \cite{pikus96, dresselhaus92, chen93}. 
This value, deduced from the data in Fig.4a and the transport parameters shown in Fig.1 , is plotted in Fig. 4b. 

When the Dresselhaus mechanism dominates  $B_{so}$ should increase with increasing density \cite{dresselhaus92} but in samples where the large spin-orbit coupling is large, such as that considered here, the Rashba term usually dominates. The Rashba  term results from structural asymmetry and is proportional to the internal electric field. Because the field is proportional to the surface charge density it should therefore increase as the concentration in the quantum well increases (see e.g. Ref. \onlinecite{Miller2003}).  
In general, the Rashba effect may therfore lead to a nonmonotonic dependence of spin-orbit splitting on gate voltage with a minimum corresponding to a symmetric quantum well. \cite{papadakis00}  In our case, however we observe a maximum of the spin orbit scattering rate (Fig. 4).  To our best knowledge this is the first report of such behavior.  
 A non-monotonic dependence of $\tau_{so}$ on electron concentration in Fig. 4 cannot be explained by either the Dresselhaus or the Rashba mechanisms and some extra effect, such as strain or the role of the interfaces must be involved. 
In the literature the role of interfaces in the Rashba mechanism is somewhat
controversial. Within the effective mass approximation the expectation value 
of a (smooth) potential gradient integrated over all space is always 
zero \cite{pikus95,gerchikov92}. More generally, the contribution from each separate interface is as large (or even larger) as that from the quantum well asymmetry \cite{gerchikov92, Grundler2000}. 
The two interfaces in a quantum well often have different properties resulting ,for example, from the growth process. 
As aresult changes in the amplitude of the electron wavefunction at each interface, produced by changes in gate voltage, will be reflected by changes in any asymmetry associated with having two different interfaces. The unexpected experimental observation that $\tau_{so}$ is a non-monotonic function of gate voltage shows such an effect plays an important role here.

An alternative method of investigating the strength of the spin-orbit coupling is to use information from the beat patterns of the low field Shubnikov-de Haas (SdH) oscillations \cite{Grundler2000, Schapers98, Dorozhkin90,Sato2001} .  There has, to our knowledge, been no published comparison of results obtained in this way with those deduced, in the same sample, from the WAL effect.  
In Ref. \onlinecite{papadakis00} the authors observed beats in SdH oscillations and an anomalous positive magnetoresistance at low field which could be due to WAL effect. However, in a more detailed study the authors \cite{papadakis02}  suggested the situation is more complex with the observed positive magnetoresistance being due to combination of several factors including the presence of  two spin subbands, a corrugated quantum well, mobility anisotropy, and "possibly weak anti-localization". 
 In the sample studied here, and also in the isomorphic sample studied previously \cite{ourWALpaper} that had a much larger spin-orbit scattering rate, a careful examination of the Shubnikov-de Haas oscillations, over a wide range of gate voltages, revealed no sign of any beats. In other samples, under conditions when there was some  parallel conduction, beats could sometimes be seen similar to those observed by other authors. However, when analyzed in detail by making gray-scale plots using many traces with small steps in gate voltage, the systematic behavior expected from spin orbit splitting could not be confirmed. In our samples we identify the beat pattern observed with interference between two sets of two SdH frequencies, originating from the gated and ungated parts of the sample, and coupled through the parasitic parallel conduction.  A similar observation has been made by Ensslin et al. \cite{Ensslin}  when they also failed to find any beats in a high quality InAs/AlSb quantum well sample.

\subsection{Conclusions}

Spin-orbit relaxation in a strained InGaAs/InP QW structure was studied using the WAL effect.  The spin-orbit relaxation time was found to depend strongly, and non-monotonically, on the gate voltage with the maximum scattering rate ($1/\tau_{so}$ = $5\times 10^{10} sec^{-1}$) reached at a density of $n=3\times 10^{15}m^{-2}$ . This behavior cannot be explained by either the Rashba or bulk Dresselhaus mechanisms but is rather attributed to asymmetry or strain effects at dissimilar QW interfaces.

Compared with a similar, but unstrained sample, the spin-orbit scattering rate here is smaller (by a factor of over 100 at the highest densities).  In the strained sample $\tau_{so}$  shows a strong gate voltage dependence (varying from 20 to 1000ps) while  in the unstrained sample $\tau_{so}$ was only weakly dependent on electron concentration.  This demonstrates that strain can be used as a tool for producing desirable spin-orbit properties when engineering materials for spintronics applications.  The exact mechanism responsible for the variation of the spin-orbit coupling in strained samples is not yet understood and is the subject of further investigations. Further theoretical work is also needed to explain correctly the experimentally observed magnetic field dependence, particularly in samples where the WAL effect is large.

\subsection{Acknowledgements}

S.A.S and A.S. acknowledge support of The Canadian Institute for Advanced Research (CIAR). 
We would like to thank S.Dickmann for helpful discussions, and J. Lapointe for assistance in  fabrication of the Hall bar.

\begin{figure}[tbp]
\includegraphics[height=70mm,clip=false]{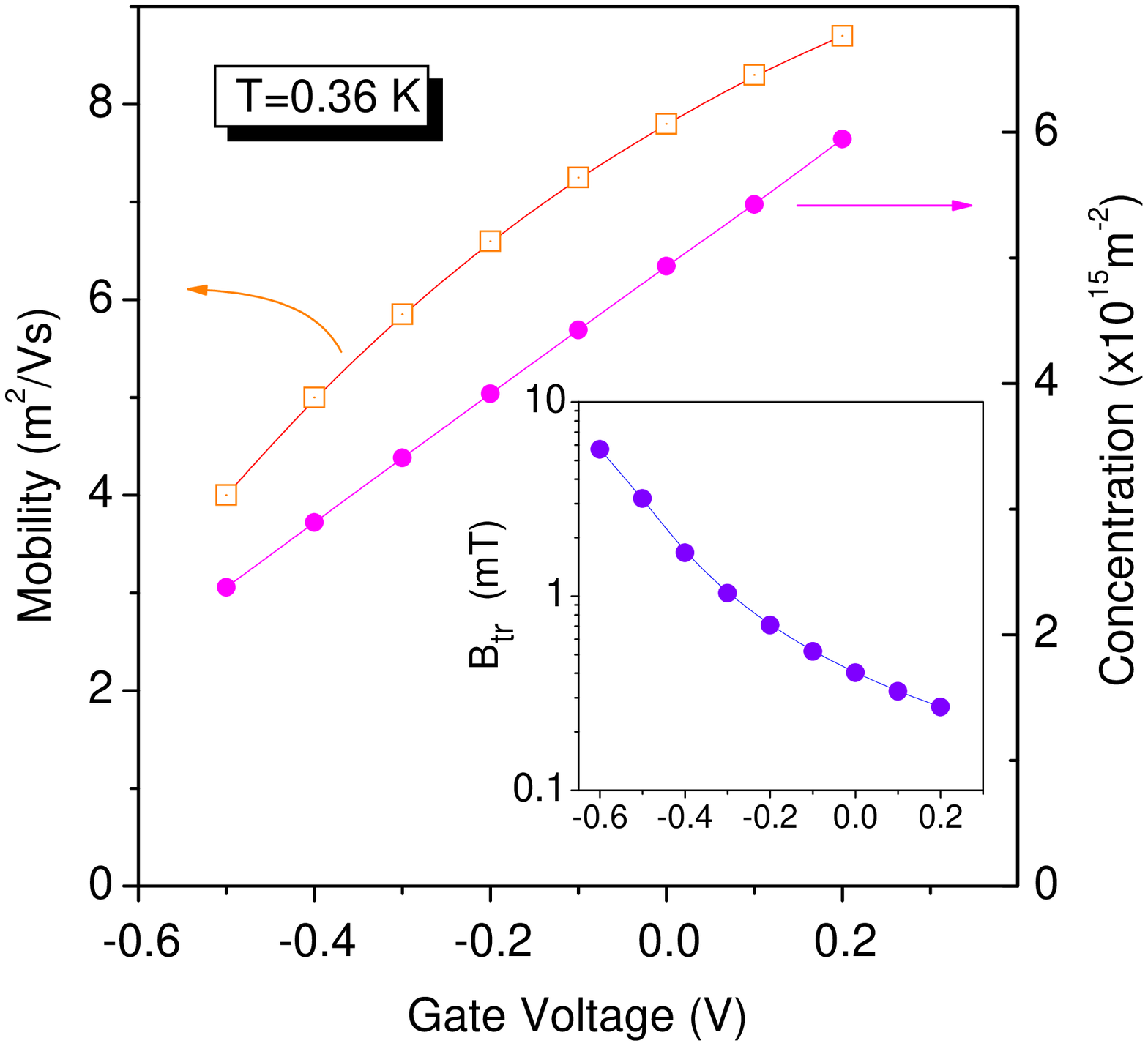}
\caption{Electron concentration (solid circles) determined from Shubnikov-de Haas oscillations and the Hall mobility (open squares) of the InGaAs/InP QW structure.  Insert shows the characteristic magnetic field  $B_{tr} = \hbar/4eD \tau$ as a function of the gate voltage. } \label{fig1}
\end{figure}

\begin{figure}[tbp]
\includegraphics[height=70mm,clip=false]{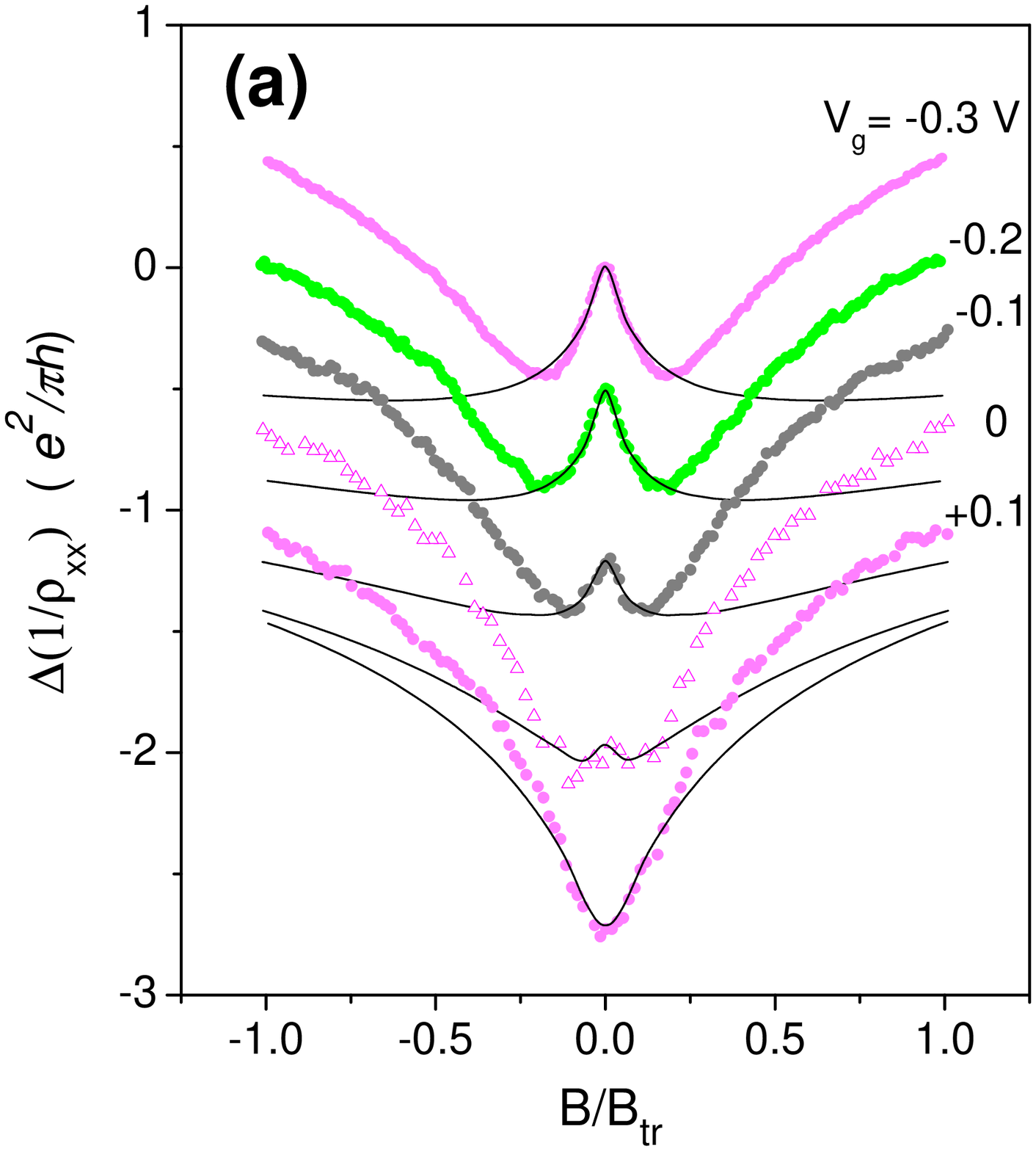}
\includegraphics[height=70mm,clip=false]{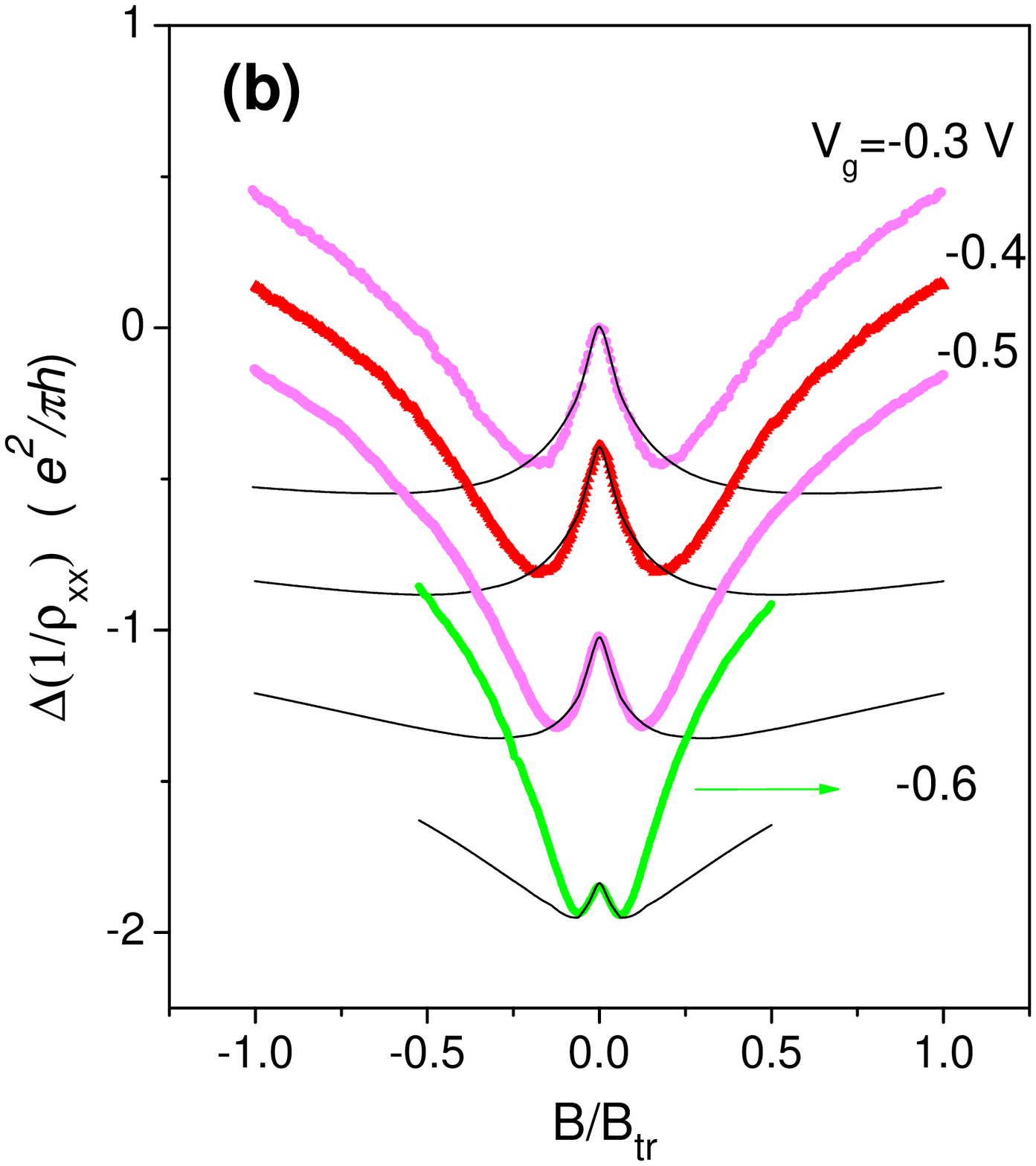}
\caption{Experimental traces of the magnetoresistance for different gate voltages at T=0.36 K. \\
(a) $V_g$= 0.1, 0, -0.1, -0.2, and -0.3 V; \\
(b) $V_g$= -0.3, -0.4, -0.5, and -0.6 V.\\
Thin solid lines are best theoretical fits to the experiment using Eq.\ref{eq1}.   } 
\label{fig2}
\end{figure}

\begin{figure}[tbp]
\includegraphics[height=70mm,clip=false]{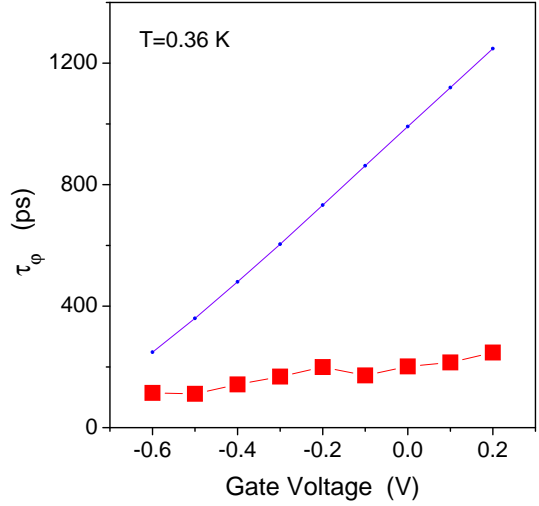}
\caption{Phase breaking time obtained from set of the data in Fig. 2  as a function of the gate voltage. Straight line is the theoretical limit due to the electron-electron scattering.   } 
\label{fig3}
\end{figure}

\begin{figure}[tbp]
\includegraphics[height=70mm,clip=false]{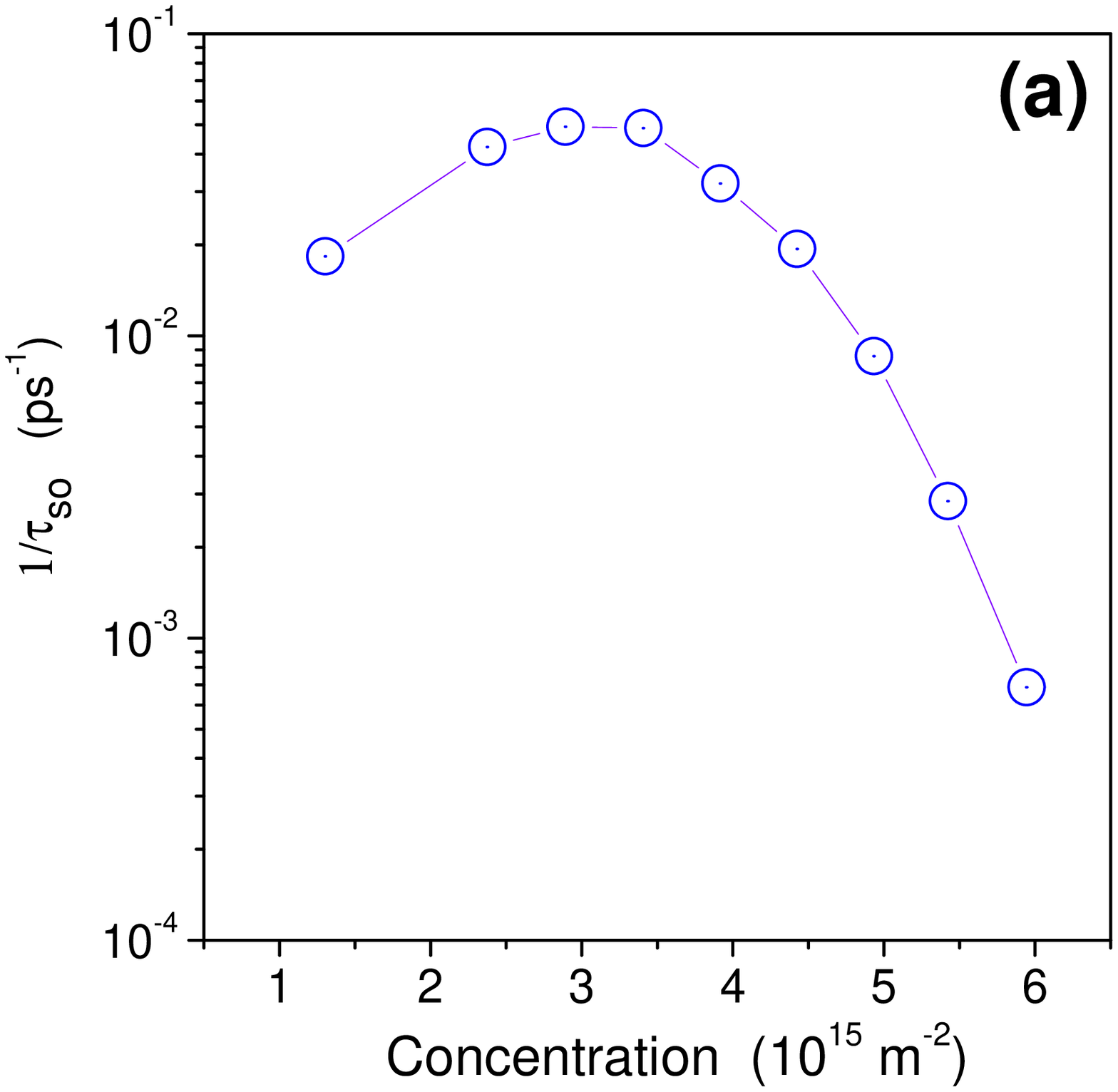}
\includegraphics[height=70mm,clip=false]{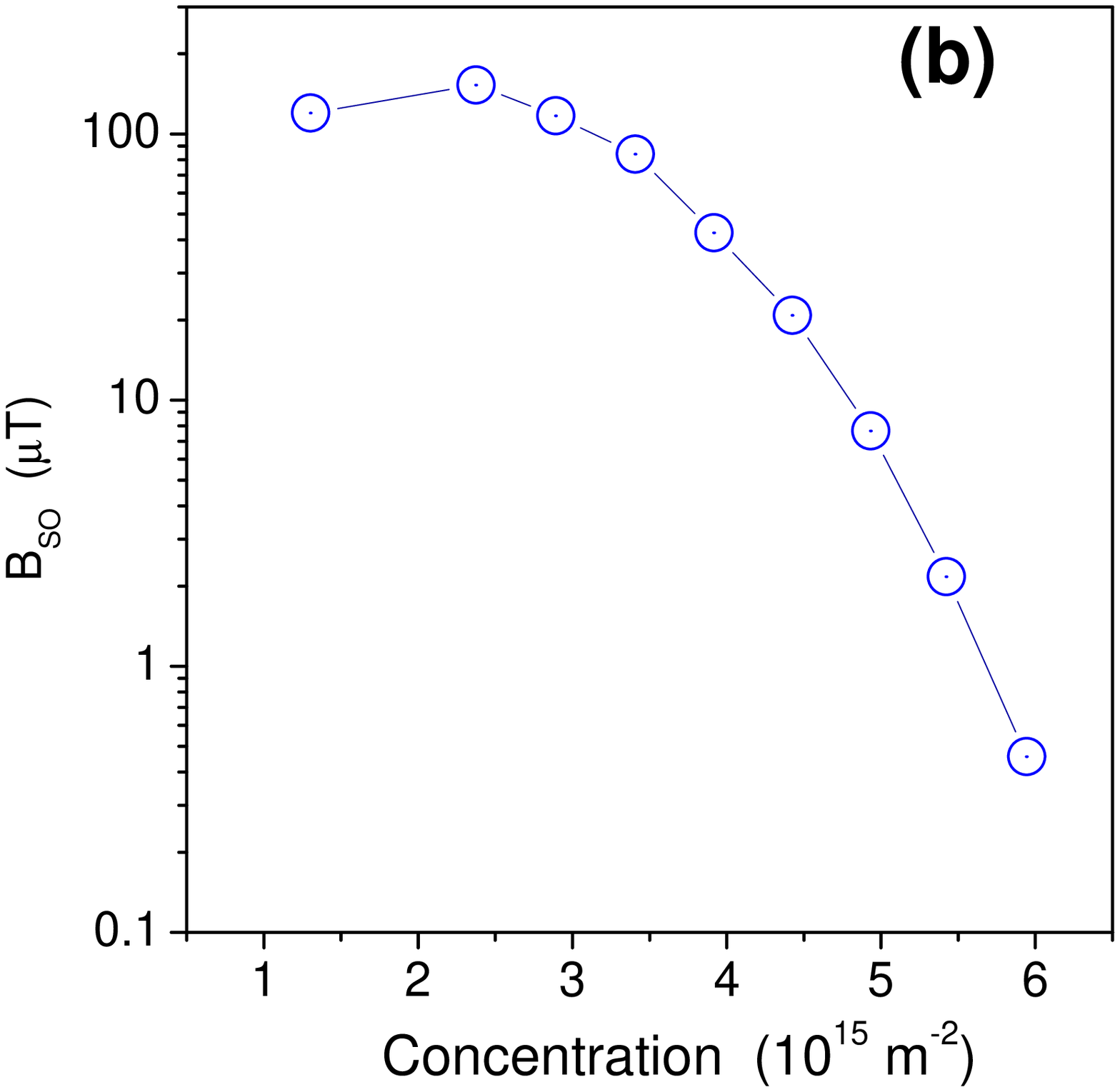}
\caption{(a) Spin-orbit scattering rate determined from  data in Fig. 2 as a function of electron concentration.\\
(b) Spin-orbit magnetic field parameter $B_{so}=\hbar/(4eD\tau_{so})$ as a function of electron concentration calculated on the basis of Figs. 1 and 4(a).    } 
\label{fig4}
\end{figure}

\end{document}